\documentclass[letterpaper]{article}

\usepackage{natbib,alifeconf}  
\usepackage{amssymb}
\usepackage{amsmath}
\usepackage{graphicx}
\usepackage{xcolor}
\usepackage{import}
\usepackage{subcaption}
\usepackage{placeins}
\usepackage[english]{babel}
\usepackage[hidelinks]{hyperref}
\usepackage[noend,noline,ruled]{algorithm2e}

\newcommand*{\ldblbrace}{\left\{\mskip-5mu\left\{}
\newcommand*{\rdblbrace}{\right\}\mskip-5mu\right\}}
\graphicspath{{figures/}}

\title{Combinatory Chemistry: \\Towards a Simple Model of Emergent Evolution}
\author{Germán Kruszewski$^{1}$\thanks{\hspace{1em}Work done while the author was at Facebook AI.}\and Tomas Mikolov$^{2}$ 
\\
\mbox{}\\
$^1$Naver Labs Europe, Grenoble, France\\
$^2$CIIRC CTU, Prague, Czech Republic\\
german.kruszewski@naverlabs.com} 

\begin{document}
\maketitle
\begin{abstract}

    An explanatory model for the emergence of evolvable units must display
    emerging structures that (1) preserve themselves in time (2) self-reproduce
    and (3) tolerate a certain amount of variation when reproducing.
    To tackle this challenge, here we introduce Combinatory Chemistry, an
    Algorithmic Artificial Chemistry based on a minimalistic computational
    paradigm named Combinatory Logic.
    The dynamics of this system comprise very few rules, it is initialized with
    an elementary \emph{tabula rasa} state, and features conservation laws 
    replicating natural resource constraints. 
    Our experiments show that a single run of this dynamical system with no
    external intervention discovers a wide range of emergent patterns.
    All these structures rely on acquiring basic constituents from the
    environment and decomposing them in a process that is remarkably similar
    to biological metabolisms.
    These patterns include autopoietic structures that maintain their
    organisation, recursive ones that grow in linear chains or binary-branching
    trees, and most notably, patterns able to reproduce themselves, duplicating
    their number at each generation.
\end{abstract}

\section{Introduction}

\begin{figure}[t]
\centering{
    \resizebox{200pt}{!}{
    \def\svgwidth{250pt}
    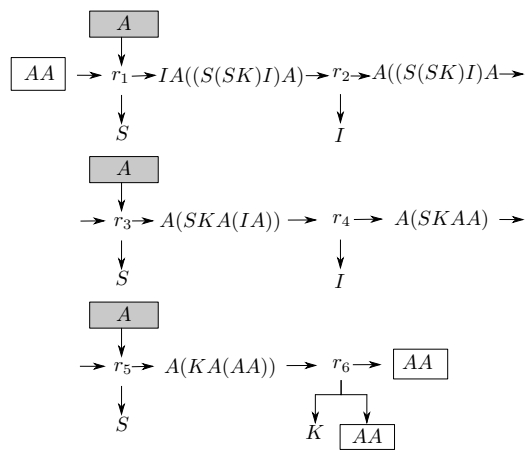
    }
}
\caption{Metabolic cycle (showing one of the possible pathways) of a self-reproducing structure that emerges from the dynamics of Combinatory Chemistry.
    Starting from $(AA)$, where $A=(SI(S(SK)I))$, it acquires three copies of
    $A$ from its environment and uses two to create a copy of itself,
    metabolising the third one to carry out the process.
}
\label{fig:self-reproducing}
\end{figure}

Finding the minimal set of conditions that lead to open-ended evolution in a
complex system is a central question in Artificial Life and a fundamental
question of science in general.
One prominent hypothesis in this line of research is that living systems
emerge from the complex interaction of simple components. 
Environments like Avida~\citep{Ofria:Wilke:2004} or Tierra~\citep{Ray:1991}
have been used to explore this question by allowing self-reproducing programs
to mutate and evolve in time.
Yet, the reproductive and mutation mechanisms, as well as the organisms' 
capacity to tolerate such mutations were fixed by design.
Instead, Artificial Chemistries try to uncover how such evolvable units emerge
in the first place by simulating the properties of natural chemical systems at
different levels of abstraction (see \citet{Dittrich:2001} for a thorough
review).
The driving hypothesis is that complex organizations emerge thanks to
self-organising attractors in chemical networks, which preserve their
structure in time \citep{Walker:1966, Wuensche:etal:1992, Kauffman:1993}.
While some Artificial Chemistries seek to mimic as closely as possible the
properties of the chemistry that gave rise to life on Earth
\citep{Flamm:etal:2010, Hogerl:2010, Young:Neshatian:2013}, others abstract
away from the particularities of natural chemistries to focus only on their
hypothesized core computational properties \citep{Fontana:1994, Speroni:2000,
Tominaga:etal:2007,Buliga:Kauffman:2014}.
In line with this latter line of work, in this paper we introduce an
Algorithmic Artificial Chemistry based on Combinatory Logic
\citep{Schonfinkel:1924, Curry:1958} featuring a minimalistic design and three
key properties.
First, it is Turing-complete, enabling it to express an arbitrary degree of
complexity.
Second, it is strongly constructive~\citep{Fontana:Buss:1993}, meaning that as
the system evolves in time it can create 
new components that can in turn modify its
global dynamics. 
Third, it features intrinsic conservation laws so that the number of atomic
elements remains always constant.

Previous work has relied on applying extrinsic conservation laws,
such as for instance keeping a maximum number of total elements in the system
by randomly removing exceeding ones~\citep{Fontana:1994, Speroni:2000}.
Instead, intrinsic conservation laws allow us to bound the total number of
elements without introducing extraneous perturbations.
Furthermore, limiting the total amount of basic elements can create selective
pressures between emergent structures.

We simulate a Chemical Reaction System ~\citep{Hordijk:etal:2015,
Dittrich:2001} based on Combinatory Logic, which starting from a \emph{tabula
rasa} state consisting of only elementary components, it produces a diversity
explosion that develops into a state dominated by self-organized emergent
structures, including autopoietic \citep{Varela:Maturana:1973}, recursive and
self-replicating ones. 
Notably, all these types of structures emerge at different points in time
during a single run of the system without requiring any external interventions.
Furthermore, these structures preserve themselves by absorbing compounds from
their environment and decomposing them step-by-step, in a process that has a
striking resemblance with the metabolism of biological organisms.
Finally, we introduce a heuristic to emulate the effects of having larger
systems without having to compute them explicitly.
This makes considerably more efficient the search for these complex
structures.

The paper is organized as follows.
First, we describe earlier work in Artificial Chemistry that is most related
to our approach.
Then, we explain the basic workings of Combinatory Logic and how we adapted it
into an Artificial Chemistry.
Third, following earlier work, we discuss how autocatalytic sets can be used to
detect emerging phenomena in this system, and propose a novel measure of
emergent complexity, which is well-adapted to the introduced system.
Finally, we describe our experiments that showcase the emergence of complex
structures in Combinatory Chemistry.


\section{Artificial Chemistries}
Artificial Chemistries (AC) are models inspired in natural chemical systems
that are usually defined by three different components: a set of possible
molecules, a set of reactions, and a reactor algorithm describing the reaction
vessel and how molecules interact with each other~\citep{Dittrich:2001}.
In the following discussion we will focus on algorithmic chemistries that are
the closest to the present work.

AlChemy~\citep{Fontana:1994} is an AC where molecules are given by
$\lambda$-calculus expressions. 
$\lambda$-calculus is a mathematical formalism that, like Turing machines, can
describe any computable function. 
In AlChemy, pairs of randomly sampled expressions are joined through function
application, and the corresponding result is added back to the population.
To keep the population size bounded, expressions are randomly discarded.
Fontana and Buss showed that expressions that computed themselves quickly
emerged in this system, which they called level 0 organisations.
Furthermore, when these expressions were explicitly prohibited, a more complex
organization emerged where every expression in a set was computed by
other expressions within the same set (level 1 organisations).
Finally, mixing level 1 organisations could lead to higher order interactions
between them (level 2 organisations).
Yet, this system had some limitations.
First, each level of organisation was only reached after external
interventions.
Also, programs must necessarily reach a normal form, which happens when there
are no more $\lambda$-calculus rules than can be applied.
Thus, recursive programs, which never reach a normal form, are banned from the
system.
Furthermore, two processes where introduced as analogues of food and waste, 
respectively.
First, when expressions are combined, they are not removed from the
system, allowing the system to temporarily grow in size.
Second, expressions which after being combined with existing expressions
do not match any $\lambda$-calculus reduction rules are removed.
Without these processes, complex organisations fail to emerge.
Yet, it is not clear under which circumstances these external interventions
would not be needed anymore in order for the system to evolve autonomously.
Finally, bounding the total number of expressions by randomly removing excess
ones creates perturbations to the system that can arbitrarily affect the
dynamics.
\cite{Fontana:Buss:1996} later proposed MC2, a chemistry based on Linear Logic
that addressed some of these limitations (notably, conservation of mass), but
we are not aware of empirical work on it.

Here, we propose an AC based on Combinatory Logic.
This formalism has been explored before in the context of AC by
\cite{Speroni:2000}.
While this work shares with us the enforcement of conservation laws, it relies
for it on a normalisation process that introduces noise into the system
dynamics.
Furthermore, as AlChemy, it reduces expressions until they reach their normal
forms, explicitly forbidding recursive and other type of expressions that do
not converge.

Finally, Chemlambda \citep{Buliga:Kauffman:2014} is a Turing-complete graph
rewriting AC that allows the encoding of $\lambda$-calculus and combinatory logic
operators. 
As such, it is complementary in many ways with the system proposed here.
Yet, we are not aware of conservation laws defined within this formalism, nor
of any reactor algorithm allowing explorations of emerging phenomena.

\section{Combinatory Logic} 
Combinatory Logic (CL) is a minimalistic computational system that was
independently invented by Moses Sch\"onfinkel, John Von Neumann and Haskell
Curry \citep{Cardone:2006}. 
Other than its relevance to computability theory, it has also been applied 
in Cognitive Science as a model for a Language of Thought \citep{Piantadosi:2016}.
One of the main advantages of CL is its formal simplicity while capturing
Turing-complete expressiveness.
In contrast to other mathematical formalisms, such as $\lambda$-calculus, it
dispenses with the notion of variables and all the necessary bookkeeping that 
comes with it.
For instance, a function $f(x) = 1 + x + y$ would be nonsensical, and a
function-generating system based on $\lambda$-calculus would need to have
explicit rules to avoid the formation of such expressions.
Instead, CL expressions are built by composing elementary operators called
combinators. 
Here, we restrict to the $S$, $K$ and $I$ combinators, which form a
Turing-complete basis\footnote{As a matter of fact, $S$ and $K$ suffice because
    $I$ can be written as $SKK$. The inclusion of $I$ simply allows to express
    more complex programs 
with shorter expressions.}. 
Given an expression $e$ of the form $e=\alpha X \beta$, it can be rewritten
in CL, as follows:
\begin{eqnarray*}
    \alpha (I f) \beta &\rhd& \alpha f \beta\\
    \alpha (K f g) \beta &\rhd& \alpha f \beta \\
    \alpha (S f g x) \beta &\rhd& \alpha (f x ( g x )) \beta
\end{eqnarray*}
When $\alpha X \beta$ matches the left hand side of
any of the rules above, the term $X$ is called a ``reducible expression'' or \emph{redex}.
A single expression can contain multiple redexes.
If no rule is matched, the expression is said to be in normal form.
The application of these rules to rewrite any redex is called a (weak)
reduction.
For example, the expression $SII(SII)$ could be reduced as follows (underlining
the corresponding redexes being rewritten):
$\underline{SII(SII)} \rhd \underline{I(SII)}(I(SII)) \rhd
SII(\underline{I(SII)}) \rhd SII(SII)$. 
Thus, this expression reduces to itself.
We will later see that expressions such as this one will be important for the
self-organizing behaviour of the system introduced here.
In contrast, $(SII)$ is not reducible because $S$ requires three arguments
and $I$ at least one.\footnote{Precedence rules are left-branching, thus $(SII) = ((SI)I)$ and thus, the second $I$ is not an argument for the first $I$ but to $SI$.}
Also note that $I(SII)(I(SII))$ has two redexes that can be rewritten, namely,
the outermost or the innermost $I$ combinators.
Even though many different evaluation order strategies have been defined
\citep{Pierce:2002}, here we opt for picking a redex at random\footnote{For
practical efficiency reasons, we restrict to sampling from the first 100
possible reductions in an outer-to-inner order.}, both because this is more
natural for a chemical system and to avoid limitations that would come from
following a fixed deterministic evaluation order.

\section{Combinatory Chemistry}

One of our main contributions deals with reformulating these reduction rules as 
reactions in a chemical system. 
For this, we postulate the existence of a multiset of CL expressions $\mathcal{P}$ that
react following reduction rules, plus random condensation and cleavages.
Note that if we were to apply plain CL rules to reduce these expressions, the
total number of combinators in the system would not be preserved.
First,
because the application of a reduction rule always removes the combinator from
the resulting expression. 
Second, while the $K$ combinator discards a part of the expression (the
argument $g$), $S$ duplicates its third argument $x$.
Thus, to make a chemical system with conservation laws, we define reduce
reactions for an expression $\alpha X \beta$, as follows:
\begin{eqnarray}
    \alpha (I f) \beta &\rightarrow& \alpha f \beta [+ I] \label{eq:red0}%
    \\
    \alpha (K f g) \beta &\rightarrow& \alpha f  \beta [+ g + K] \label{eq:redK} %
    \\
    \alpha (S f g x) \beta \underbrace{[+ x]}_{\text{reactant}} &\rightarrow& \alpha (f x (g x)) \beta \underbrace{[+ S]}_{\text{by-product}}\label{eq:redS}
    \label{eq:redn}
\end{eqnarray}
An expression in Combinatory Chemistry is said to be \emph{reducible} if it
contains a Combinatory Chemistry redex (CC-redex).
A CC-redex is a plain CL redex, except when it involves the reduction of an $S$
combinator, in which case a copy of its third argument $x$ (the reactant) must
also be present in the multiset $\mathcal{P}$ for it to be a redex in
Combinatory Chemistry.
For example, the expression $SII(SII)$ is reducible if and only if the third
argument of the combinator $S$, namely $(SII)$, is also present in the set.
When a reduction operation is applied, the redex is rewritten following the
rules of combinatory logic, removing any reactant from $\mathcal{P}$ and adding
back to it all by-products, as specified in brackets on the right hand size of
the reaction.
The type of combinator being reduced gives name to the reaction.
For instance, the $S$-reaction operating on $SII(SII) + (SII)$ removes these
two elements from $\mathcal{P}$, adding back $I(SII)(I(SII))$ and $S$ to it.
Notably, each of these reduction rules preserves the total number of
combinators in the multiset, intrinsically enforcing conservation laws in this
chemistry.
It is also worth noticing that each of these combinators plays different roles
in the creation of novel compounds. 
While $K$-reactions split the expression, decreasing its total size and
complexity, $S$-reactions create larger and possibly more complex expressions
from smaller parts.

Completing the set of possible reactions in this chemistry, condensations
and cleaveges can generate novel expressions through random recombination:
\begin{eqnarray}
    x + y \longleftrightarrow xy \label{eq:random}
\end{eqnarray}

In Combinatory Chemistry, computation takes precedence.
This means that whenever an expression admits at least one reduce reaction,
this reaction (or one at random, if there are multiple) is immediately applied.
Otherwise, if an expression cannot be reduced, it is either cleaved at a random
point, or condensed together with another irreducible expression.
As reducing reactions take priority over random recombination ones, we 
construe them as auto-catalysed.

In line with the Gillespie algorithm~\citep{Gillespie:1977}, the system is
simulated by sampling expressions from $\mathcal{P}$ with probability
proportional to their concentration and applying one reaction at a time.
In this way, we uniformly distribute the computational budget between all
programs in the system.
Moreover, we do not need to take additional precautions to avoid recursive
expressions that never reach a normal form, allowing these interesting
functions to form part of our system's dynamics.

Finally, we note that while other chemistries start from a population of
randomly constructed compounds, this system can be initialised with elementary
combinators only.
In this way, diversity materializes only as emergent property rather than
through the product of an external intervention.

The complete algorithm describing the temporal evolution of our system is
summarized on Algorithm \ref{al:reactor}\footnote{\label{code}Implementation
and supplementary materials at \url{https://germank.github.io/combinatory-chemistry}}.

\newcommand{\smalltt}[1]{\texttt{\footnotesize #1}}\
\begin{algorithm}[ht]
    \SetInd{0em}{1.1em}
    \DontPrintSemicolon
    \KwIn{Total number of combinators $N_I$, $N_K$, $N_S$}
    \SetKwFunction{IsReducible}{Reducible?}
    \SetKwFunction{Reductions}{Reductions}
    \SetKwData{Combine}{combine}
    \SetKwData{Break}{break}
    \SetFuncSty{smalltt}
    Initialize multiset $\mathcal{P} \gets \ldblbrace I: N_I , K: N_K , S: N_S \rdblbrace$\;
    \While{True}{
        Sample $e \in \mathcal{P}$ with ${P}(e) = \frac{\mathcal{P}[e]}{|\mathcal{P}|}$ \;
        \If{\IsReducible{$e$}}{
            Let $\left(e [+ x] \rightarrow \hat{e} [+ y]\right) \in \Reductions{e}$\;
            Remove one $e$ from $\mathcal{P}$\;
            Remove one reactant $x$ from $\mathcal{P}$ (if applicable)\;
            Add one $\hat{e}$ and all by-products $y$ to $\mathcal{P}$\;
        }
        \Else {
            Randomly pick {\it cleave} or {\it condense}\;
            \If {cleave} {
                Let $x,y$ such that $e = (xy)$\;
                Remove one $e$ from $\mathcal{P}$\;
                Add one $x$ and one $y$ to $\mathcal{P}$\;
            }
            \ElseIf {$e_{\text{\sc left}}$ \text{\normalfont is defined}}{
                Remove one $e$ and one $e_\text{\sc left}$ from $\mathcal{P}$\;
                Add $(e_{\text{\sc left}} e)$ to $\mathcal{P}$\;
                Undefine $e_\text{\sc left}$\;
            }
            \Else {
                Define $e_{\text{\sc left}} \gets e$\;
            }
        }
    }
    \SetKwProg{Fn}{Function}{:}{}
    \SetKw{KwAnd}{ and }
    \SetKw{KwOr}{or}
    \SetKw{KwAny}{any}
    \SetKw{KwExists}{exists}
    \SetKw{KwTrue}{True}
    \SetKw{KwRet}{return}
    \SetKw{KwFalse}{False}
    \Fn{\IsReducible{$e$}}{
        \If {$e = (I f)$ \KwOr $e = (K f g)$ \KwOr \\
            $\big(e = (S f g x)$ \KwAnd $x \in \mathcal{P}\big)$}{
            \KwRet \KwTrue\;
        }
        \ElseIf{$e \in \{S, K, I\}$}{ 
            \KwRet \KwFalse\;
        }
        \Else{
            \KwRet $\exists e' : e = (\alpha{}e'\beta) \KwAnd \IsReducible{e'}$
        }
    }
    \caption{Reactor Algorithm}\label{al:reactor}
\end{algorithm}

\section{Emergent Structures}

Having described the dynamics of Combinatory Chemistry, we now turn to discuss
how can we characterise emergent structures in this system. 
For this, we first discuss how can autocatalytic sets be applied for this
purpose.
Second, we observe that this formalism may not completely account for some
emergent structures of interest and thus, we propose to instead track reactant
consumption rates as a proxy metric to uncover the presence of these structures.
Finally, inspired by the concept of food sets in autocatalytic sets, we propose
a heuristic to accelerate their emergence.

\subsection{Autocatalytic sets}

Self-organized order in complex systems is hypothesized to be driven by the
existence of attracting states in the system's dynamics~\citep{Walker:1966,
Wuensche:etal:1992, Kauffman:1993}.
Autocatalytic sets~\citep{Kauffman:1993} were first introduced by Stuart
Kauffman in 1971 as one type of such attractors that could help explaining the
emergence of life in chemical networks.
(See \citet{Hordijk:2019} for a comprehensive historical review on the topic.)
Related notions are the concept of autopoiesis \citep{Varela:Maturana:1973},
and the hypercycle model \citep{Eigen:1978}.

Autocatalytic sets (AS) are reaction networks that perpetuate in time by
relying on a network of catalysed reactions, where each reactant and catalyst
of a reaction is either produced by at least some reaction in the network, or
it is freely available in the environment.
This notion was later formalized in mathematical form
\citep{Hordijk:Steel:2004,Hordijk:etal:2015} with the name of Reflexively
Autocatalytic Food-generated sets (RAFs).
Specifically, they define a Chemical Reaction System (CRS) as a mathematical
construct defining the set of possible molecules, the set of possible reactions
and a catalysis set indicating which reactions are catalysed by which
molecules.
Furthermore, a set of freely available molecules in the environment, called the
“food set”, is assumed to exist.
An autocatalytic set (or RAF set) $\mathcal{S}$ of a CRS with associated food
set $F$ is a subset of reactions, which is:

\begin{enumerate}
\item \emph{reflexively autocatalytic} (RA): each reaction $r \in \mathcal{S}$
is catalysed by at least one molecule that is either present in $F$ or can be
formed from $F$ by using a series of reactions in $\mathcal{S}$ itself. \label{it:ra}
\item \emph{food-generated} (F): each reactant of each reaction in
$\mathcal{S}$ is either present in $F$ or can be formed by using a series of
reactions from $\mathcal{S}$ itself. \label{it:f}
\end{enumerate}

\subsection{Autocatalytic sets in Combinatory Chemistry}

\begin{figure}[t]
\centering{
    \resizebox{240pt}{!}{
    \def\svgwidth{350pt}
    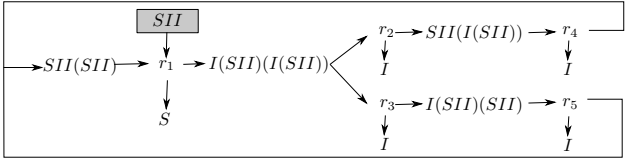
    }
}
\caption{$r_1$--$r_5$ form an autocatalytic
set, granted that $(SII)$ belongs to the food set. $(SII(SII))$'s metabolic cycle
starts with $r_1$ reducing the $S$ combinator, while taking $(SII)$ as
reactant.
Then, the cycle is completed by the reduction of the two identity combinators,
in any of the possible orders.} 
\label{fig:autocatalytic_set}
\end{figure}

In Combinatory Chemistry, all reducing reactions take precedence over random
condensations and cleavages, and thus, they proceed at a higher rate than random
reactions without the need of any catalyst (i.e. they are auto-catalysed).
Therefore, they trivially satisfy condition \ref{it:ra}.
Thus, autocatalytic sets in this system are defined in terms of subsets of
reduce reactions in which every reactant is produced by a reduce reaction in
the set or is freely available in the environment (condition \ref{it:f}).
For example, if we assume that $A=(SII)$ is in the food set, Figure
\ref{fig:autocatalytic_set} shows a simple emergent autocatalytic set
associated with the expression $(AA)=(SII(SII))$.
As shown, a chain of reduce reactions keep the expression in a self-sustaining
equilibrium:
When the formula is first reduced by reaction $r_1$, a reactant $A$ is absorbed
from the environment and one $S$ combinator is released.
Over the following steps, two $I$ combinators are sequentially applied and
released back into the multiset $\mathcal{P}$, with the expression returning
back to its original form.
We refer to this process as a metabolic cycle because of its strong resemblance
to its natural counterpart.
For convenience, we write this cycle as $(AA) + A \twoheadrightarrow (AA) +
\phi(A)$, where $\phi$ is a function that returns the atomic combinators in $A$ and
the double head arrow means that there exists a pathway of reduction reactions
from the reactives in the left hand side to the products in the right hand
side.

\begin{figure}[htb]
    \centering
    \resizebox{170pt}{!}{
    \def\svgwidth{240pt}
    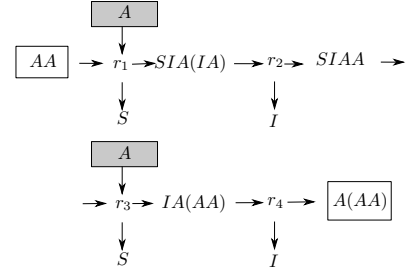
    }
\caption{One of the possible pathways in the reduction of the tail-recursive structure
$(AA)$ with $A=(S(SI)I)$. It appends one $A$ to itself by metabolising another copy
absorbed from the environment.} 
\label{fig:recursive}
\end{figure}

While autocatalytic sets provide a compelling formalism to study emergent
organization in Artificial Chemistries, it also leaves some blind spots for
detecting emergent structures of interest.
Such is the case for \textbf{recursively growing expressions}.
Consider, for instance, $e = (S(SI)I(S(SI)I))$. 
This expression is composed of two copies of $A=(S(SI)I)$ applied to itself
$(AA)$. 
As shown in Figure \ref{fig:recursive}, during its metabolic cycle it will
consume two copies of the element $A$, metabolising one to perform its
computation, and appending the other one to itself, thus $(AA)  +
2A\twoheadrightarrow (A(AA)) + \phi(A)$.
As time proceeds, the same computation will take place recursively, thus
$(A(AA)) + 2A \twoheadrightarrow (A(A(AA))) + \phi(A)$, and so on.
While this particular behaviour cannot be detected through autocatalytic sets,
because the resulting expression is not exactly equal to the original one,
it still involves a structure that preserves in time its functionality.

Moreover, while the concept of autocatalytic set captures both patterns that
perpetuate themselves in time and patterns that also multiply their numbers, it
does not explicitly differentiate between them.
A pattern with a metabolic cycle of the form $AA + A \twoheadrightarrow AA +
\phi(A)$ (as in Figure \ref{fig:autocatalytic_set}) keeps its own structure in
time by metabolising one $A$ in the food set, but it does not self-reproduce. 
We call such patterns \textbf{simple autopoietic}~\citep{Varela:Maturana:1973}.
In contrast, for a pattern to be \textbf{self-reproducing} it must create
copies of itself that are later released as new expressions in the environment.
For instance, consider a metabolic cycle in Figure \ref{fig:self-reproducing}
with the form $(AA) + 3A \twoheadrightarrow 2(AA) + \phi(A)$.
This structure creates a copy of itself from 2 freely available units of $A$
and metabolises a third one to carry out the process.

All these structures have in common the need to absorb reactants from
the environment to preserve themselves in homoeostasis.
Furthermore, because they follow a cyclical process, they will continually 
consume the same types of reactants.
Thus, we propose tracking \emph{reactants consumption} as a metric that can
capture all these different types of structures.
For this, we note that the only operation that allows an expression to
incorporate a reactant into its own body is the reduction of the $S$
combinator, and thus focus on only counting the reactants consumed by
$S$-reactions.

\subsection{Reactant assemblage}

We also note that emergent structures must necessarily rely on freely available
expressions that are at least produced by the environment through random
collisions.
However, longer reactants come in exponentially smaller concentrations, and
thus exponentially larger systems should be simulated for them to arise in
large numbers.
This makes the experimental process considerably inefficient, particularly for
allowing the emergence of complex structures that depend on such reactants.
Here, we introduce a heuristic that we call \emph{reactant assemblage}
to facilitate the exploration of larger systems without needing to simulate
them in full. 
The central idea is to arbitrarily define a food set containing the expressions
that would be freely available in a larger system. 
For this, we fix a maximum food size $F$. 
Then, whenever an $S$-reduction requires a reactant that is not present in
$\mathcal{P}$, but is part of this predefined food set, the reactant would be
constructed on the spot from freely available atomic combinators.
More precisely, we modify Algorithm \ref{al:reactor} at the point of sampling
a reduction with the steps in Algorithm \ref{al:cooking}.
In this way, we can simulate the productivity of sufficiently large
environments, without explicitly needing to compute them.
However, this technique does not bypass the need of discovering the
substrate, namely, the expression $e$ being reduced.
Instead, it just focuses on creating the required reactants.
Furthermore, as the total number of combinators in the system is limited, 
ceiling effects can be observed if the number of freely available atoms start
to dwindle.
Even with these concerns in mind, we experimentally show that this heuristic
facilitates the emergence of more complex patterns.

\begin{algorithm}[ht]
    \DontPrintSemicolon
    \KwIn{Maximum reactant size $F$}
    \SetKw{KwAnd}{and}
    \SetKwFunction{Combinators}{Combinators}
    Let $\left(e [+ x] \rightarrow \hat{e} [+ y]\right) \in \text{\Reductions{e}}$\; 
    \If{$x \not\in \mathcal{P}$ \KwAnd $|x| \leq F$}
    {
        Let $n_I, n_K, n_S = \Combinators{x}$\;
        \If{$\mathcal{P}[I] \geq n_I$ \KwAnd $\mathcal{P}[K] \geq n_K$ \KwAnd $\mathcal{P}[S] \geq n_S$}{
            Remove $n_k$ $K$ from $\mathcal{P}$\;
            Remove $n_I$ $I$ from $\mathcal{P}$\;
            Remove $n_S$ $S$ from $\mathcal{P}$\;
            Add one $x$ to $\mathcal{P}$\;
        }
    }
    \caption{Reactant assemblage}\label{al:cooking}
\end{algorithm}

\section{Experiments and Discussion}

\begin{figure*}[hbt]
    \centering
    \begin{subfigure}[t]{0.3\textwidth}
        \includegraphics[width=\linewidth]{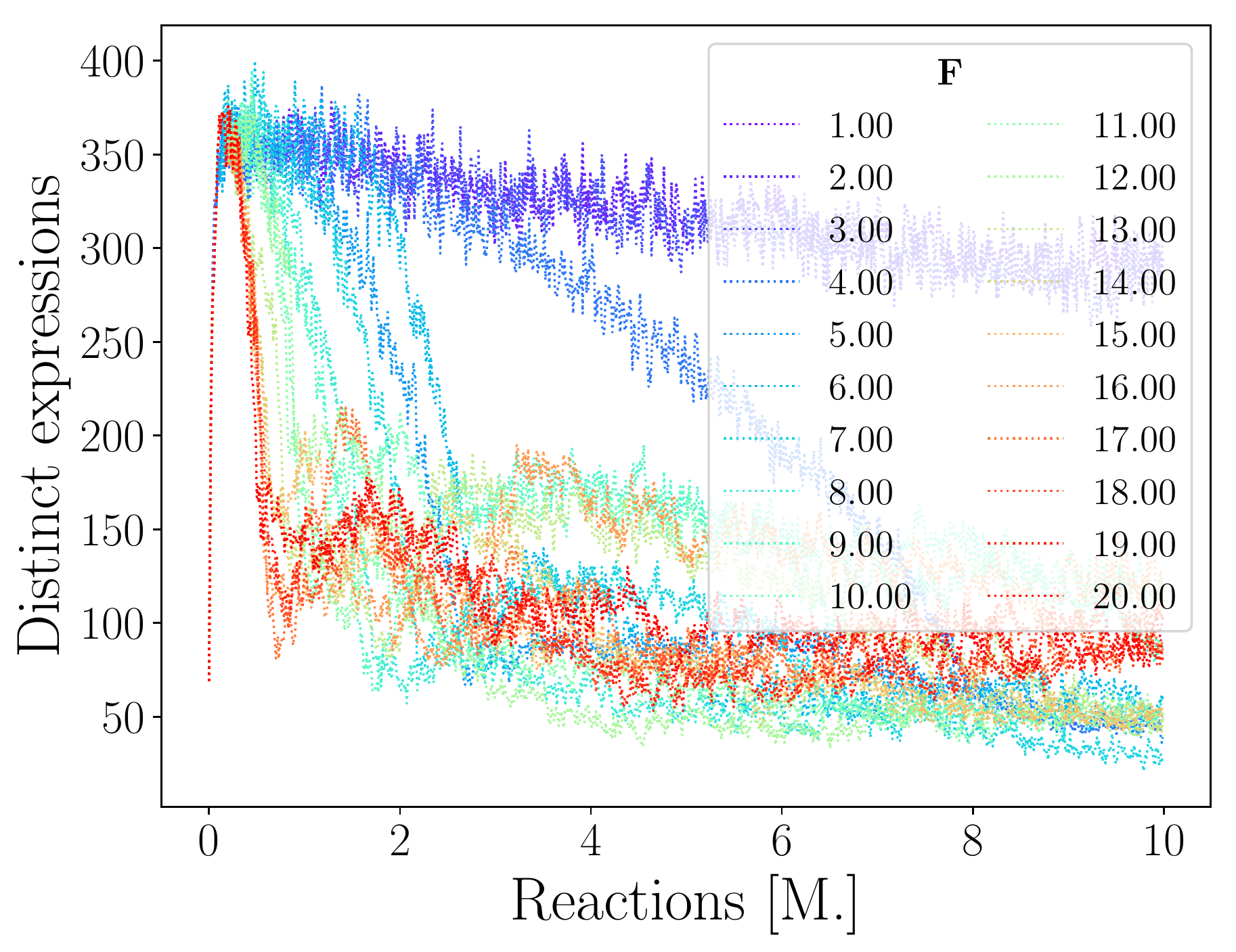}
        \caption{Diversity}
        \label{fig:diversity}
    \end{subfigure}
    \begin{subfigure}[t]{0.3\textwidth}
        \includegraphics[width=\linewidth]{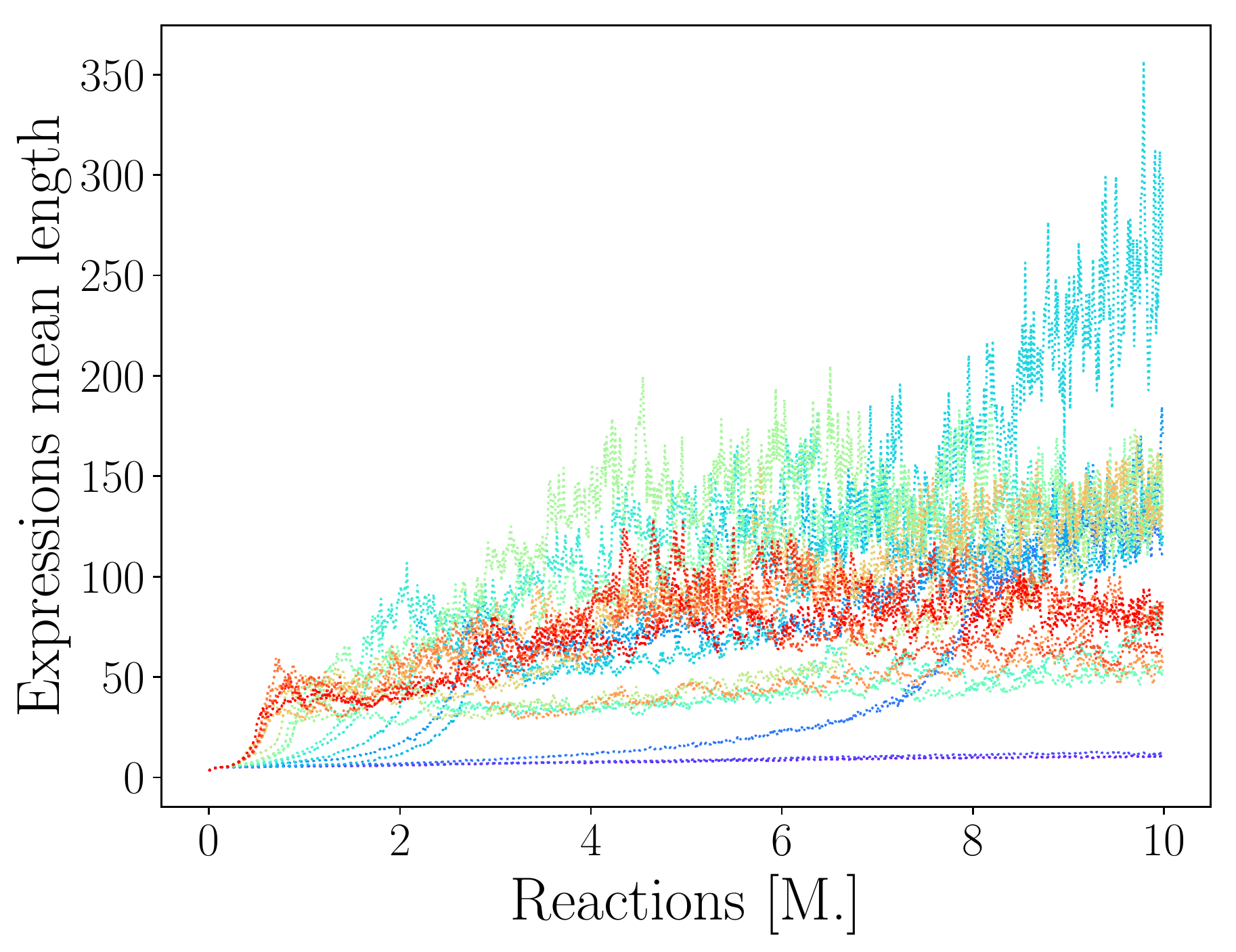}
        \caption{Mean expressions length}
        \label{fig:length}
    \end{subfigure}
    \begin{subfigure}[t]{0.3\textwidth}
        \includegraphics[width=\linewidth]{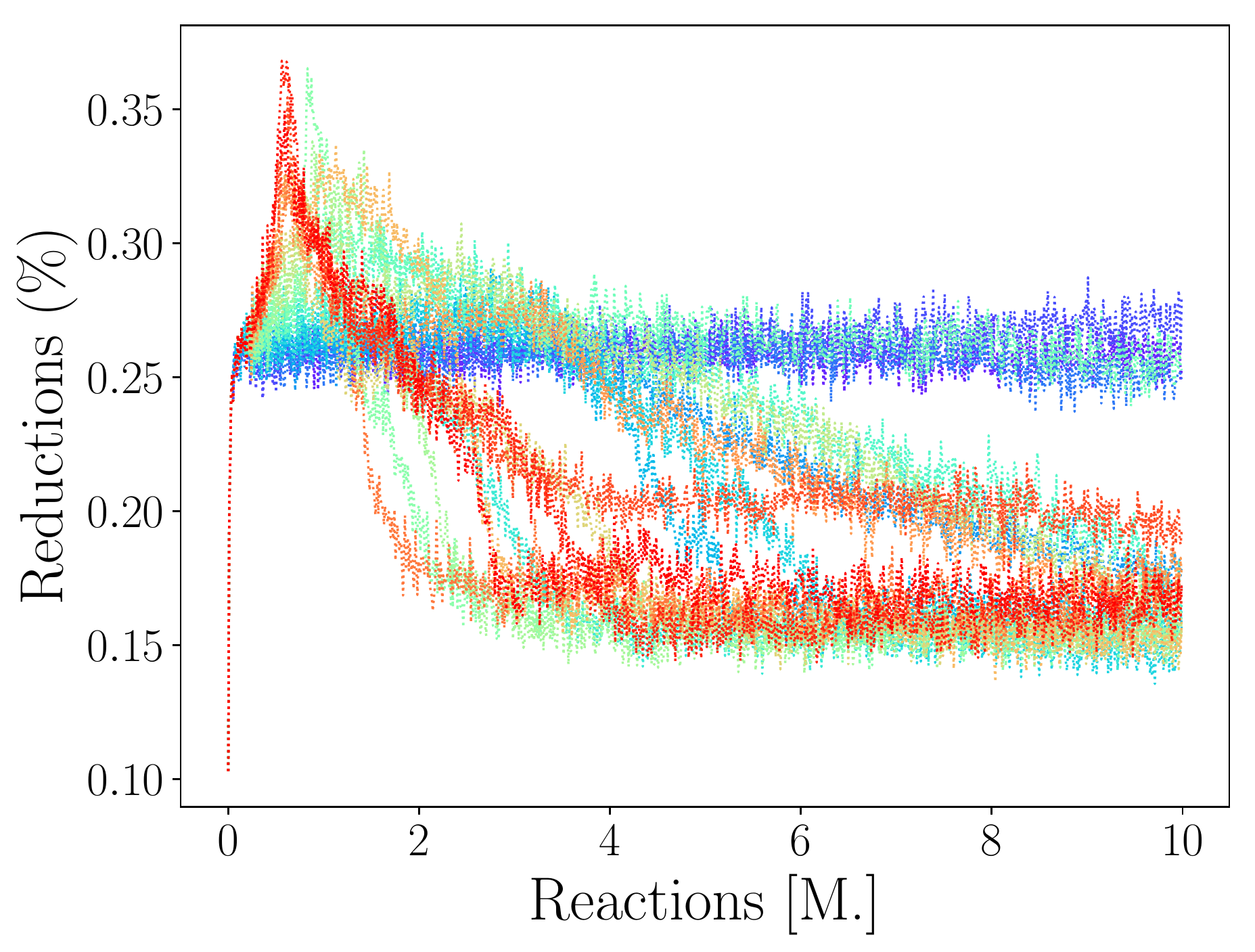}
        \caption{Reductions}
        \label{fig:p_reduce}
    \end{subfigure}
    \caption{System metrics for different values of reactant assemblage size $F$.}
\end{figure*}

We initialized $\mathcal{P}$ with $10$k evenly distributed $S$, $K$, and $I$
combinators and applied reactant assemblage with reactant size parameter $F$ on
a range between $1$ (corresponding to no reactant assemblage) and $20$, 
simulating $10$ different runs of Combinatory Chemistry for $10$M iterations.

We then began by analysing general metrics of the system for different values
of $F$.
Figure \ref{fig:diversity} shows the expression diversity as a function of the
number of performed reactions. 
As it can be seen, diversity explodes in the first few $200$k reactions, before
reaching a peak of about $300$ different expressions. 
Then, it starts to decline at different speeds, depending on the value of $F$.
When this mechanism is disabled ($F=1$), the decline occurs at a slow and
steady rate.
Yet, when $F=3$, the decline of diversity becomes much faster, only
accelerating with higher $F$.
This effect could be explained by the fact that $S$-reductions, the only
ones that compute increasingly longer expressions, are more likely to be
successful thanks to the reactant assemblage mechanism getting into action.
Therefore, the limited available combinators tend to be clustered in fewer and
longer expressions. 
This is also consistent with the evolution of the mean expression length shown
in Figure \ref{fig:length}.
Yet, when $F$ is set to the high end of the range, reduce operations peak, but
then start being replaced by random cleavages and condensations as freely
available combinators needed to assemble reactants plummet (Figure
\ref{fig:p_reduce}).
Thus, the system self-regulates the ratio of deterministic operations (reductions)
to random ones (condensations and cleavages) occurring in it.
\begin{figure*}[t]
    \captionsetup[subfigure]{justification=centering}
    \centering
    \begin{subfigure}[t]{0.3\textwidth}
        \centering
        \includegraphics[width=\linewidth]{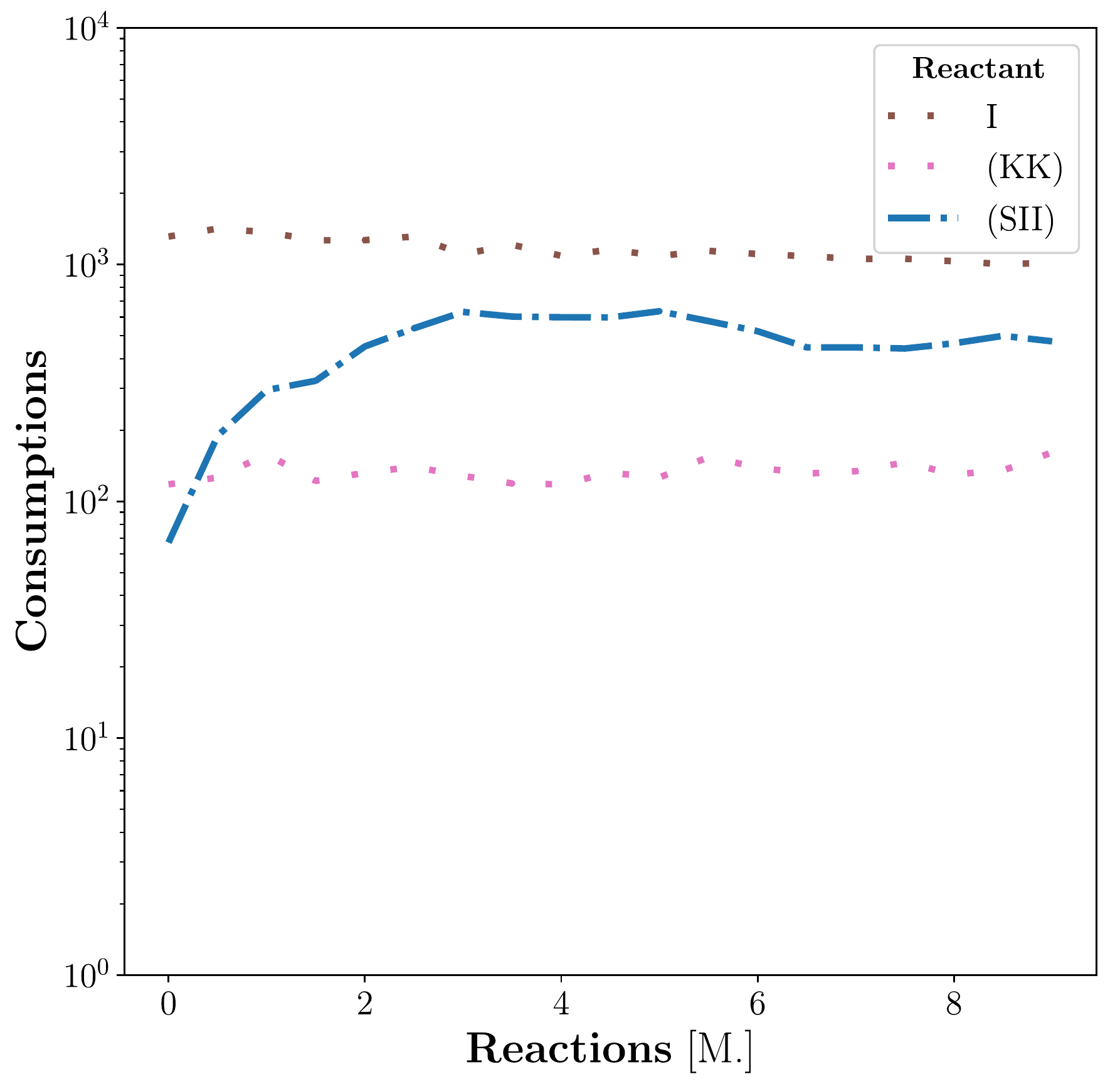}
        \caption{$F=1$}
        \label{fig:sub_1}
    \end{subfigure}%
    ~ 
    \begin{subfigure}[t]{0.3\textwidth}
        \centering
        \includegraphics[width=\linewidth]{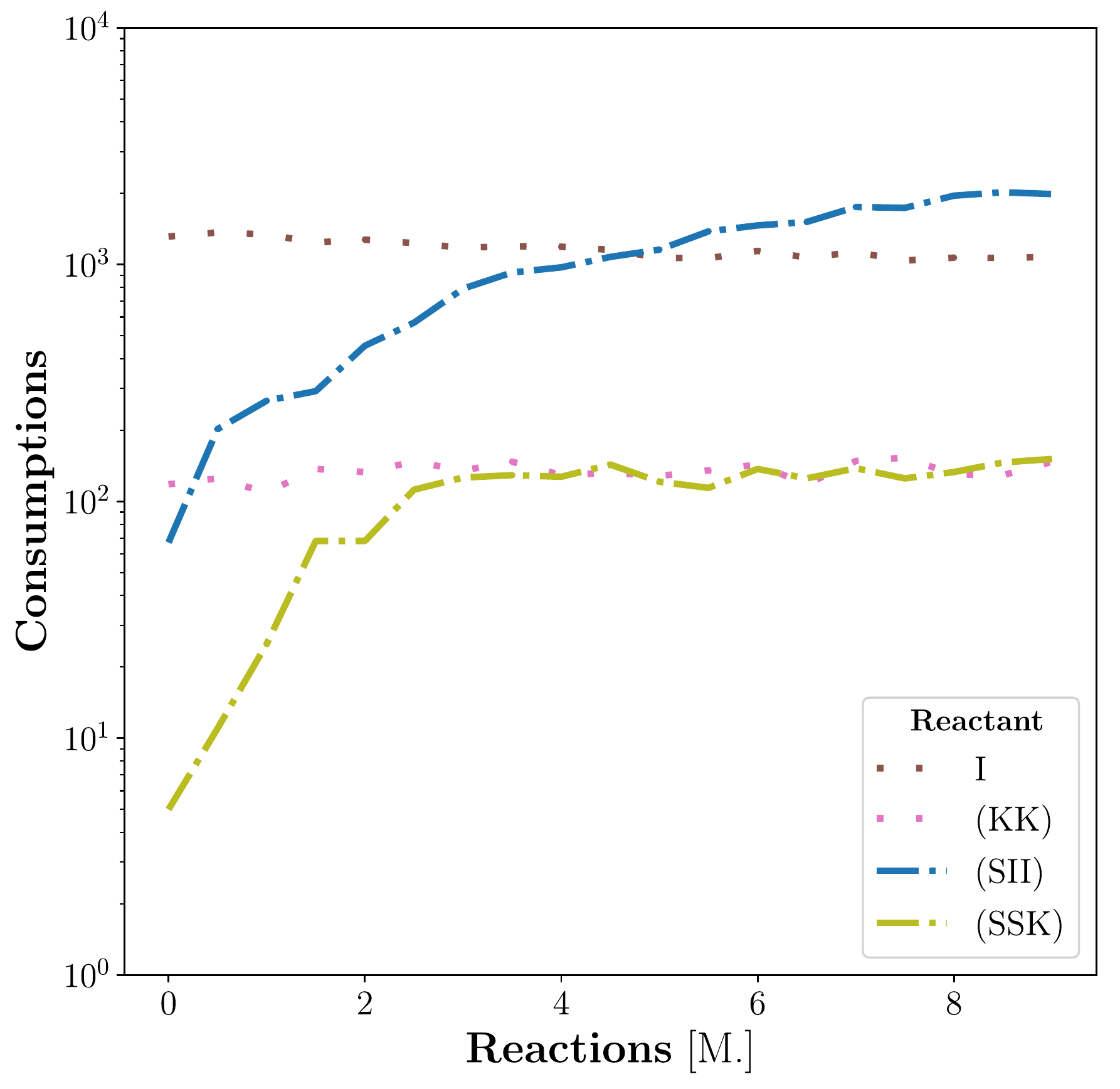}
        \caption{$F=3$}
        \label{fig:sub_3}
    \end{subfigure}
    \begin{subfigure}[t]{0.3\textwidth}
        \centering
        \includegraphics[width=\linewidth]{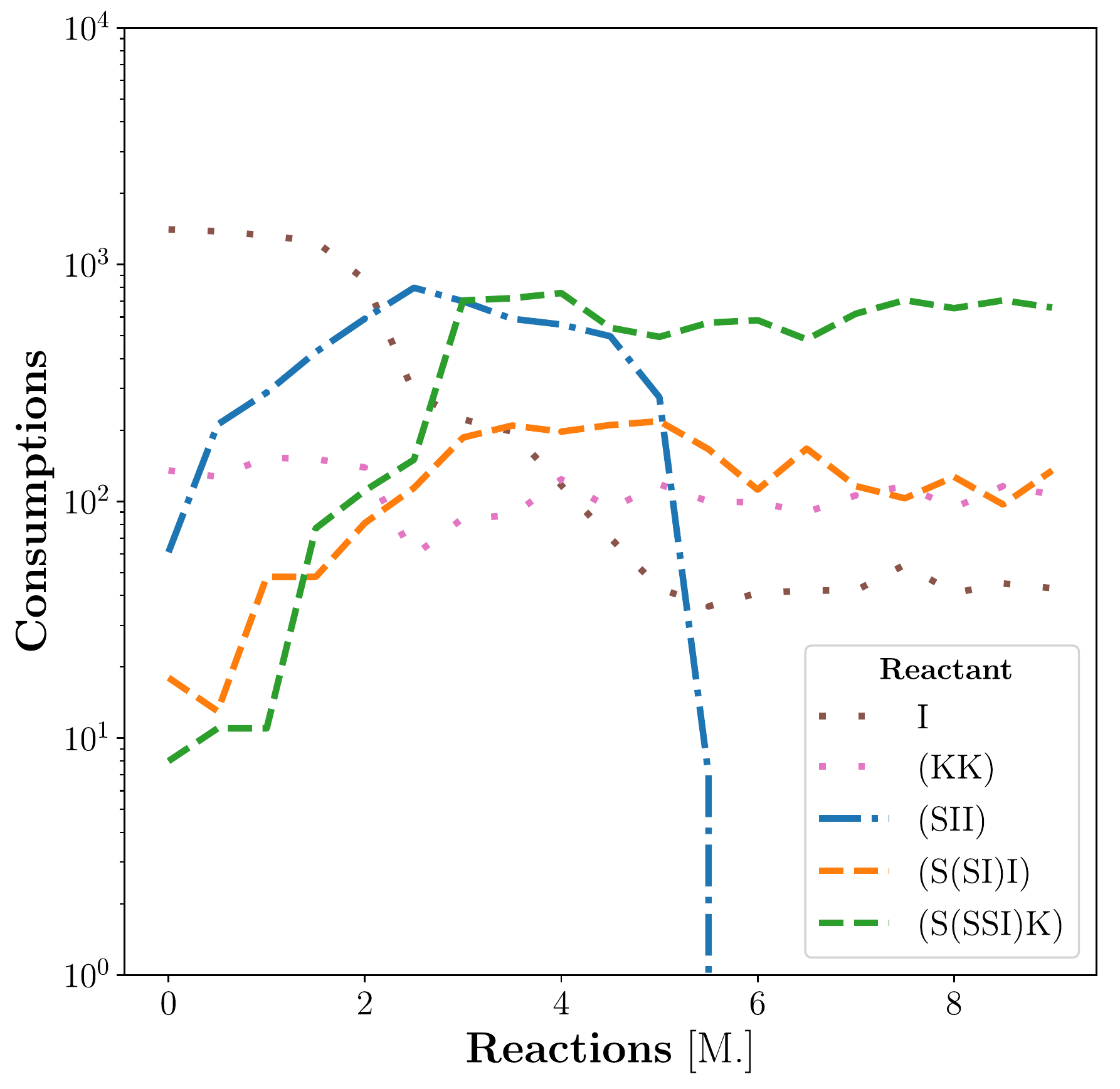}
        \caption{$F=6$}
        \label{fig:sub_6}
    \end{subfigure}
    \begin{subfigure}[t]{0.3\textwidth}
        \centering
        \includegraphics[width=\linewidth]{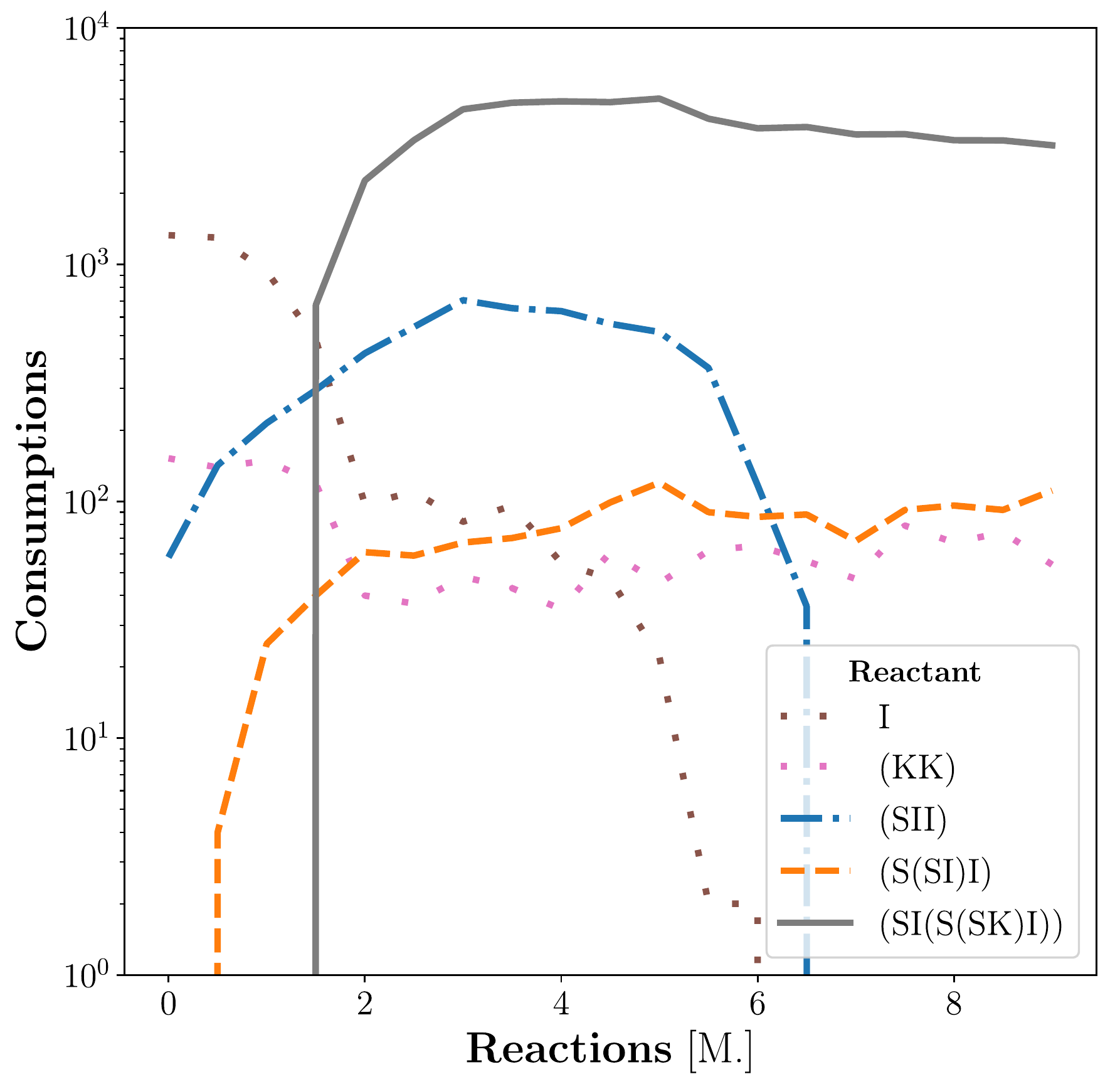}
        \caption{$F=6$}
        \label{fig:sub_6b}
    \end{subfigure}
    \begin{subfigure}[t]{0.3\textwidth}
        \centering
        \includegraphics[width=\linewidth]{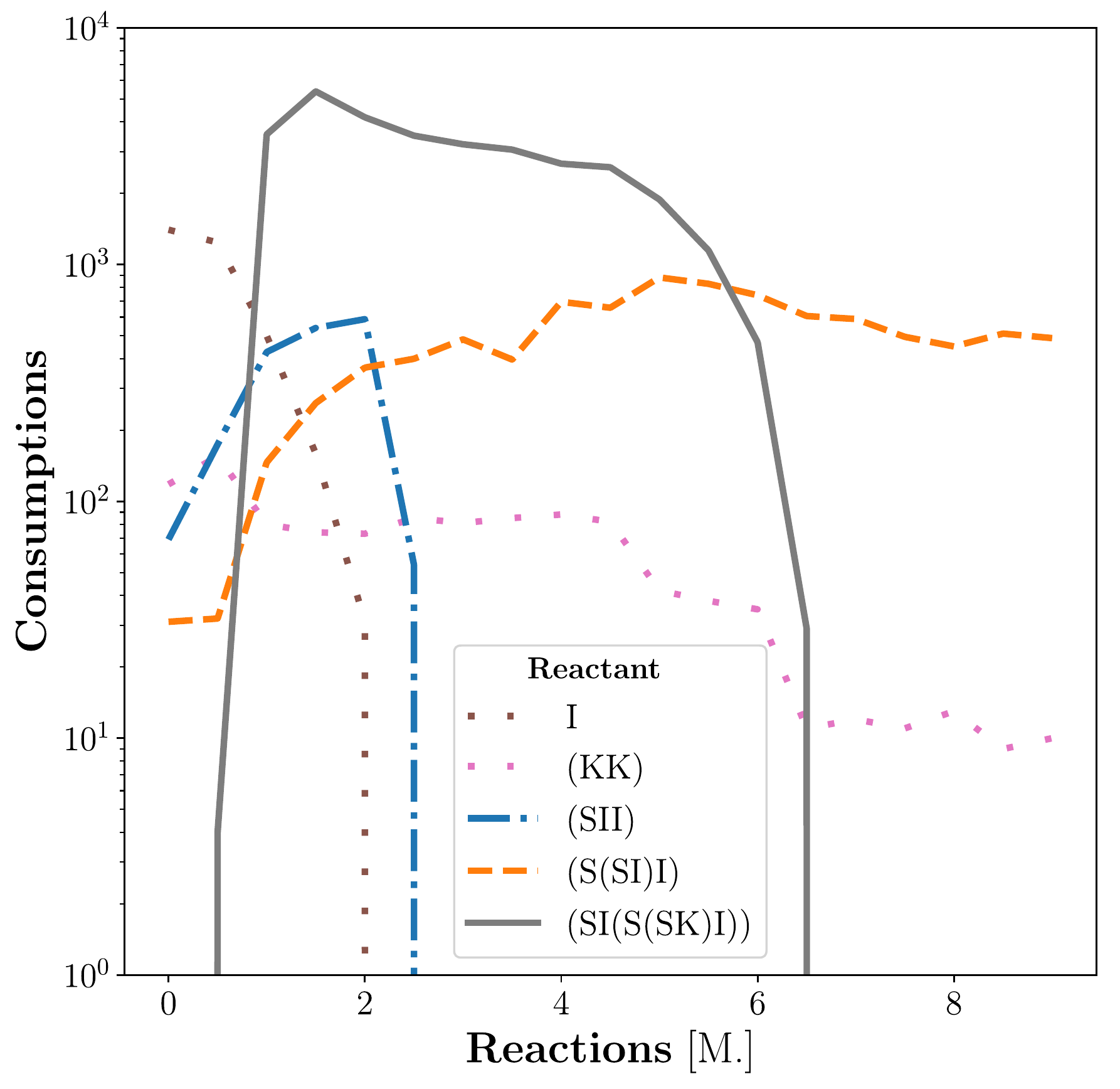}
        \caption{$F=8$}
        \label{fig:sub_8}
    \end{subfigure}%
    \begin{subfigure}[t]{0.3\linewidth}
        \centering
        \includegraphics[width=\linewidth]{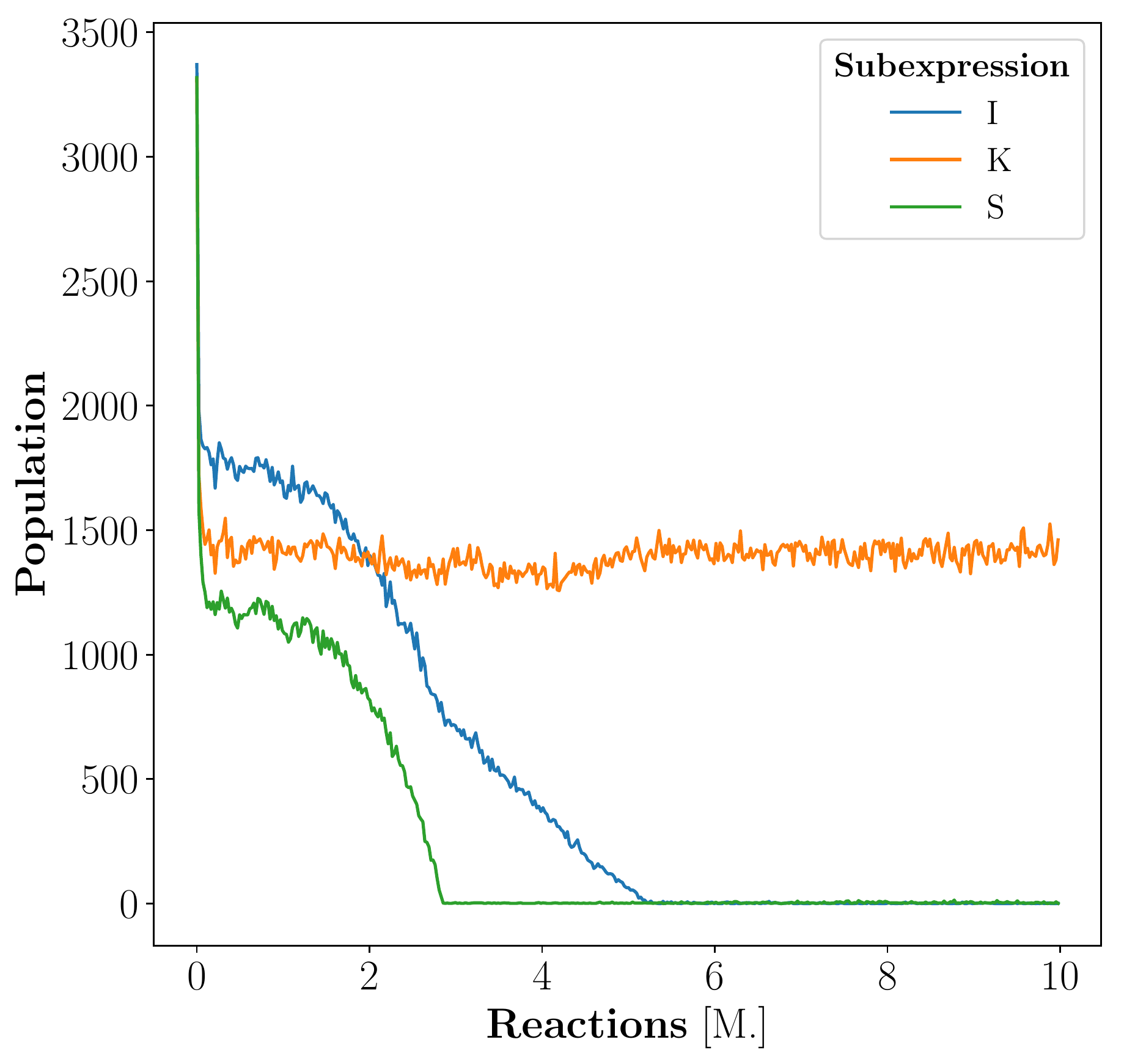}
        \caption{Combinator population in the simulation of Fig. (c)}
        \label{fig:ext}
    \end{subfigure}
    \caption{(a-e) Reactants consumption computed over a window of $500$k reactions
    on different runs with different reactant assemblage sizes $F$.
    Different line styles distinguish reactants used by different types
    of expressions:
    Dotted lines represent reactants consumed by simple non-attracting expressions;
    dash-dotted lines are reactants used by simple \emph{autopoietic} patterns;
    dashed lines are reactants used by \emph{recursively growing} patterns;
    solid lines are reactants used by \emph{self-reproducing} patterns.
    }
    \label{fig:reactants}.
\end{figure*}

%
However, it is unclear from these results whether the ensuing reduction in
diversity is driven by emergent complex structures that act as attractors, or
by some other different reason.
To answer this question, we tracked reactant consumption rates to detect
whether specific reactants where more prominently used.
In particular, we looked at the 10 most frequently used reactants, from which
we are selecting a few to simplify the presentation.
Results are shown in Figure \ref{fig:reactants} for five different runs.
In each of these, we can see the emergence of different types of structures,
including simple autopoietic, recursively growing, and self-reproducing
patterns.
Interestingly, they can emerge at different points in time, co-exist, and
sometimes some of them can drive others to extinction.

We first observed that expressions that consume any given reactant $A$ are
typically composed of multiple juxtaposed copies of this reactant, confirming
the old adage: ``Tell me what you eat and I will tell you what you are''.
For instance, in Figure \ref{fig:sub_1} we can appreciate the emergence of the
autopoietic pattern $(SII(SII))$, composed of two copies of the reactant
$A=(SII)$, and a metabolic cycle of the form $(AA) + A \twoheadrightarrow (AA)
+ \phi(A)$, as shown on Figure \ref{fig:autocatalytic_set}.

Binary reactants such as $(KK)$ and unitary ones such as $I$ do not form
part of any stable structure, and the expressions consuming them are produced
by chance.
Yet, they are used with considerable frequency because $S$ combinators are
more likely to be applied to shorter arguments than longer ones.
For this reason, the consumption of $I$ is considerable higher than the
consumption of $(KK)$.
Yet, even though by the same argument the consumption rate of $A=(SII)$ should
be below binary reactants, self-organization into autopoietic patterns drives
the usage of this reactant above what would be expected would chance be the
only force at play.
At around 2M-3M reactions, the system reaches a point in which the consumption
levels of this reactant stabilizes, constrained by the free availability of 
the reactant.
Yet, when we assemble reactants of size at most $F=3$ (Figure \ref{fig:sub_3}),
the availability of $SII$ reactants is greatly expanded as long as $S$ and $I$
combinators are freely available in the environment, thus allowing the
formation of an even larger number of $(SII(SII))$ structures.

At the same time, rarer autopoietic structures can emerge in this condition,
such as one based on three juxtaposed copies $(AAA)$ of $A=(SSK)$. 
This expression has a metabolic cycle of the form $(AAA) + 2A
\twoheadrightarrow (AAA) + A + \phi(A)$\footnote{Derivations are available at
the supplementary materials.}.
Here, we note that one copy of $A$ is used but then released intact, which could be
construed as an emergent catalyst for the reaction:
Indeed, even though we can interpret each reduce reaction to be auto-catalysed,
reaction chains can have emergent properties, such as in this
case, where a reactant is just used to complete the metabolic cycle and then
released.

From $F=4$ we start to see growing structures.
In particular, \emph{recursively growing} ones. 
In Figure \ref{fig:sub_6} ($F=6$) we can observe two such structures.
The first one uses the reactant $A=(S(SI)I)$, and follows a tail-recursive
cycle that linearly increases the size of the structure: $(AA) + 2A
\twoheadrightarrow A(AA) + \phi(A)$ (Figure \ref{fig:recursive}).
The second one is a more complex binary-branching recursive structure, with
reactant $A=(S(SSI)K)$.
When recursive structures come into play, we can see that simple autopoietic
patterns are driven into extinction.
These extinction events are related to the assemblage mechanism, as it puts
recursive and autopoietic structures in direct competition for atomic combinators. 
Figure \ref{fig:ext} displays the amount of freely available combinators
in this simulation.
As shown, $S$ combinators are exhausted at around 3M reactions.
At around this point the simple autopoietic structure consuming the $(SII)$
reactant goes into a slow decline.
Yet, when all freely available $I$ combinators are depleted, this structure is
driven into a quick extinction:
Without the needed reactant $S$-reactions start to fail, and thus the
expression is either cleaved or combined with another one.
When $(SII(SII))$ is broken into two independent $(SII)$ elements it loses its
ability to compute itself.
%
In contrast, recursive structures can cope with conditions of low resources
quite effectively, as demonstrated by the fact that they still continue to
consume at stable rate their corresponding reactants after $S$ and $I$
combinators are not freely available anymore.
A possible reason why this does not bring them into catastrophic failure is
their fractal structure:
A recursive structure broken up will still have the same function, but it
will be smaller.
For instance, $A(AA) \rightarrow A + (AA)$ still leaves a functioning $(AA)$
structure.
When new resources become available through the continuous influx of
combinators released by every computed reduction, it can consume them and grow
back again.
In the future, to avoid such direct competition for basic resources between all
emergent structures, the reactant assemblage could be limited, for instance,
to be applied only when there is a minimum buffer of freely available
combinators.

Finally, in Figures \ref{fig:sub_6b} and \ref{fig:sub_8} we can observe the
emergence of a full-fledged \emph{self-reproducing} structure with reactant
$A=(SI(S(SK)I))$. 
It follows a cycle of the form $(AA)  + 3A \twoheadrightarrow 2(AA) + \phi(A)$,
thus duplicating itself, and metabolising one reactant in the process.
As it replicates exponentially, this structure quickly grows into one of the
most active ones. 
Yet, when resources run out it enters in direct competition with the recursive
structure based on the $(S(SI)I)$ reactant.
In \ref{fig:sub_6b} the reactant consumption rate for this last structure is
considerably lower than in \ref{fig:sub_8}, seizing less resources for itself.
This may be due to the fact that the recursive structure $(A(AA))$ can
either reduce the internal part, consuming one copy of $A$, or at the most
external level, consuming $(AA)$, which in the latter case is facilitated by
reactant assemblage when $F=8$.
Yet, the self-reproducing pattern suffers from the same problem of simple
autopoietic structures: When it fails to acquire its reactant from the
environment it decomposes into an expression that loses its functionality.
However, in contrast with simple autopoietic patterns that rely on being
produced by chance, self-reproducing ones can recover their population
through reproduction. 
Nevertheless, the recursive structure still keeps an advantage over the
self-reproducing one, especially when $F=8$, where it quickly drives the
self-reproducing pattern into extinction.

\section{Conclusions}

We have introduced Combinatory Chemistry, an Algorithmic Artificial Chemistry
based on Combinatory Logic. 
Even though it has simple dynamics, it gives rise to a wide range of complex
structures, including recursively growing and self-reproducing ones.
Thanks to Combinatory Logic being Turing-complete, the presented system can
theoretically represent patterns of arbitrary complexity.
Furthermore, the computation is distributed uniformly across the system thanks
to single-step reactions applied at each iteration while conservation laws keep
the system bounded without introducing any extrinsic perturbations.
The emerging structures that result from these dynamics feature reaction cycles
that bear a striking resemblance to natural metabolisms.

Moreover, this system does not need to start from a random set of initial
expressions to kick-start diversity.
Instead, this initial diversity is the product of the system's own dynamics, as
it is only initialized with elementary combinators.
In this way, we can expect that this first burst of diversity is not just a 
one-off event, but it is deeply embedded into the mechanics of the system, 
possibly allowing it to keep on developing novel structures continually.

Finally, we noted that emerging structures require a constant influx of
specific types of reactants.
While only much larger systems than the ones simulated here would allow for
such continual production of many of these food elements, we proposed a
heuristic to make them available without explicitly simulating them.
In this way, have observed a wide variety of emerging structures, including
those that would self-sustain, albeit not changing their numbers (simple
autopoietic); recursive expressions that would keep growing until reaching the
system's limit; and self-reproducing patterns that increase their number
exponentially.

To conclude, we have introduced a simple model of emergent complexity in which
self-reproduction emerges autonomously from the system's own dynamics.
In the future, we will seek to apply it to explaining the emergence of
evolvability, one of the central questions in Artificial Life.
While many challenges lie ahead, we believe that the simplicity of the model,
the encouraging results presently obtained, and the creativity obtained from
balancing computation with random recombination to search for new forms,
leaves it in good standing to tackle this challenge.

\section{Acknowledgements}

We thank the anonymous reviewers and Mattia De Stefano for their feedback,
Sebastian Riedel for his invaluable support, and the FAIR
London team, the Complexity Interest Group, and the CSSS'19 Machine Creativity
reading group for the interesting discussions. 
A special thanks to Alessandra Macillo for suggesting the idea behind reactant
assemblage and for all the feedback, and Jordi Piñero for his literary
recommendations.

\footnotesize
\bibliographystyle{apalike}
\bibliography{german} 

\end{document}


\maketitle
\renewcommand{\thesection}{\Alph{section}}
\section{Appendix}
\label{ap:derivations}

The following derivations show one of the possible pathways that each of the
described structures can undertake as they develop.
They  demonstrate how can these structures preserve their function in time.
Note that every expression written as $(fX)Y$ can also be written simply as
$fXY$. This ``trick'' is known as Uncurrying.

\subsection{Metabolic cycle of a simple autopoietic pattern}
Let $A=(SII)$. Then,

\begin{eqnarray*}
(\underline{AA}) + A &\rightarrow& ((IA)(IA)) + S\\
(\underline{IA}(IA)) & \rightarrow & (A(IA)) + I \\
(A(\underline{IA})) & \rightarrow & (AA) + I \\
\end{eqnarray*}

\subsection{Metabolic cycle of a ternary autopoietic pattern}

Let $A=(SSK)$. Then, 

\begin{eqnarray*}
(\underline{AA}A) + A &\rightarrow& (SA(KA)A) + S\\
(\underline{SA(KA)A}) + A & \rightarrow & (AA(KAA)) + S\\
(AA(\underline{KAA})) & \rightarrow & (AAA) + A + K \\
\end{eqnarray*}

\subsection{Metabolic cycle of a tail recursive structure}
Let $A=(S(SI)I)$. Then,

\begin{eqnarray*}
(\underline{AA}) + A &\rightarrow& (SIA(IA)) + S\\
(SIA(\underline{IA})) & \rightarrow & SIAA + I\\
(\underline{SIAA}) + A & \rightarrow & (IA(AA)) + S\\
(IA(AA)) & \rightarrow & (A(AA)) + I \\
\end{eqnarray*}

\subsection{Metabolic cycle of a binary-branching structure}

Let $A=(S(SSI)K)$. Then $(AA)$ can follow the metabolic pathway:

\begin{eqnarray*}
(\underline{AA}) + A &\rightarrow& (SSIA(KA)) + S\\
\underline{SSIA}(KA) + A&\rightarrow& (SA(IA)(KA)) + S\\
(SA(\underline{IA})(KA)) &\rightarrow& (SAA(KA)) + I\\
(\underline{SAA(KA)}) + (KA) &\rightarrow& (A(KA)(A(KA))) + S
\end{eqnarray*}

Then each copy of $(A(KA))$ can be reduced as follows

\begin{eqnarray*}
(\underline{A(KA)}) + (KA) &\rightarrow& SSI(KA)(K(KA)) + S\\
(\underline{SSI(KA)}(K(KA)) + (KA) &\rightarrow& (S(KA)(I(KA))(K(KA))) + S\\
(S(KA)(\underline{I(KA)})(K(KA))) &\rightarrow& (S(KA)(KA)(K(KA))) + I\\
(\underline{S(KA)(KA)(K(KA))}) + (K(KA)) &\rightarrow& (KA(K(KA))(KA(K(KA)))) + S\\
(\underline{KA(K(KA))}(KA(K(KA)))) &\rightarrow& (A(KA(K(KA)))) + (K(KA)) + K\\
(A(\underline{KA(K(KA))})) &\rightarrow& (AA) + (K(KA)) + K
\end{eqnarray*}

Thus, the complete pathway can be summarized as $(AA) +2A + 5(KA) + (K(KA))
\twoheadrightarrow AA(AA) + 4 (K(KA)) + 2\phi(A)$.

\subsection{Metabolic cycle of a self-reproducing expression}

Let $A=(SI(S(SK)I))$

Then,

\begin{eqnarray*}
(\underline{AA}) + A &\rightarrow& (IA(S(SK)IA)) + S \\
(\underline{IA}(S(SK)IA)) &\rightarrow& (A(S(SK)IA)) + I\\
(A(\underline{S(SK)IA}))  + A & \rightarrow & (A(SKA(IA))) + S\\
(A(SKA(\underline{IA}))) & \rightarrow & (A(SKAA)) + I \\
(A(\underline{SKAA})) & \rightarrow & (A(KA(AA))) + S \\
(A(\underline{KA(AA)})) & \rightarrow & (AA) + (AA) + K \\
\end{eqnarray*}